\documentclass{article}
\usepackage{ijcai19}
\usepackage{times}
\usepackage{soul}
\usepackage{url}
\usepackage{xcolor}
\usepackage[hidelinks]{hyperref}
\usepackage[utf8]{inputenc}
\usepackage[small]{caption}
\usepackage{graphicx}
\usepackage{amsmath}
\usepackage{booktabs}
\usepackage{algorithm}
\usepackage{algorithmic}
\usepackage{multirow}
\usepackage{float}

\title{Alzheimer’s Disease Brain MRI Classification: Challenges and Insights}

\author{
Yi Ren Fung\thanks{The authors contributed equally} \and
Ziqiang Guan\footnotemark[1] \and
Ritesh Kumar\and
Joie Yeahuay Wu\And
Madalina Fiterau
\affiliations
College of Information and Computer Sciences\\
University of Massachusetts Amherst\\
\emails
\{yfung, zguan, riteshkumar, yeahuaywu, mfiterau\}@cs.umass.edu
}

\begin{document}

\maketitle

\begin{abstract}
In recent years, many papers have reported state-of-the-art performance on Alzheimer's Disease classification with MRI scans from the Alzheimer's Disease Neuroimaging Initiative (ADNI) dataset using convolutional neural networks. However, we discover that when we split the data into training and testing sets at the subject level, we are not able to obtain similar performance, bringing the validity of many of the previous studies into question. Furthermore, we point out that previous works use different subsets of the ADNI data, making comparison across similar works tricky. In this study, we present the results of three splitting methods, discuss the motivations behind their validity, and report our results using all of the available subjects.

\end{abstract}

\section{Introduction}

Alzheimer's Disease is a progressive neurodegenerative disease characterized by cognitive decline and memory loss. It is one of the leading causes of death in the aging population, affecting millions of people around the world. Automatic diagnosis and early detection can help identify high risk cohorts for better treatment planning and improved quality of life \cite{alzheimer2017facts}. 

Initial works on brain MRI classification rely on domain-specific medical knowledge to compute volumetric segmentation of the brain in order to analyze key components that are strongly correlated with the disease \cite{gerardin2009multidimensional,plant2010automated,bhagwat2018modeling}. With recent advances in deep learning and an increasing number of MRI scans being made public by initiatives such as Alzheimer's Disease Neuroimaging Initiative (adni.loni.usc.edu) and Open Access Series of Imaging Studies (www.oasis-brains.org), various convolutional neural network (CNN) architectures, including 2D and 3D variants of ResNet, DenseNet, InceptionNet, and VGG models, have been proposed to holistically learn feature representations of the brain \cite{jain2019convolutional,khvostikov20183d,korolev2017residual,wang2018AD}. A patient's diagnosis in the dataset is typically categorized as Alzheimer's disease (AD); mild cognitive impairment (MCI); or cognitively normal (CN). 

While many previous related works report high accuracy in three-class classification of brain MRI scans from the ADNI dataset, we point out a major issue that puts into question the validity of their results. A number of papers \cite{payan2015predicting,farooq2017deep,wu2018discrimination,jain2019convolutional} perform classification with training and testing splits done randomly across the brain MRIs, in which the implicit assumption is that the MRIs are independently and identically distributed. As a result, a given patient's scans from different visits could exist across both the training and testing sets. 

We point out that transitions in disease stage occur at a very low frequency in the ADNI dataset, which means that a model can overfit to the brain structure corresponding to an individual patient and not learn the key differences in feature representation among the different disease stages. The model can simply regurgitate patient-level information learned during training and rely on that to obtain good classification results on the test set. We refer to generating the splits by sampling at random from the pool of images, ignoring the patient IDs, as \textit{splitting by MRI randomly}.

Our main contribution in this study is exploring alternative ways to split the data into training and testing sets such that the evaluation gives valid insights into the performance of the proposed methods. 

First, we look at \textit{splitting the data by patient}, where all of the available visits of a patient are assigned to either the training or testing set, with no patient data split across the two sets. This method is motivated by the use case for automatic diagnosis from MRI scans, where the assumption is that no prior or subsequent patient information is given.

Then, we look at \textit{splitting the data by visit history}, meaning that the first $n-1$ visits of a patient's data are used for training, and the $n$\textsuperscript{th} visit is used for testing. This method is motivated by the use case wherein we incorporate MRI scans into personalized trajectory prediction.

Finally, we compare the accuracy of training and testing on sets split randomly by MRI, with the accuracies we obtain using the two aforementioned ways of splitting the training and testing set. We run experiments with both popular 2D models pre-trained on ImageNet, as well as a 22-layer ResNet-based 3D model we built. Unlike many other studies that perform experiments on a subset of the ADNI dataset, our experiments include all of the currently available data from ADNI. Our results suggest that there is still much room for improvement in classifying MRI scans for Alzheimer's Disease detection. 

\section{Related Work}

Before the current wave of interest in deep learning, the field of neuroimaging relied on domain knowledge to compute some predefined sets of handcrafted metrics on brain MRI scans that reveal details on Alzheimer's disease progression, such as the amount of tissue shrinkage that has occurred. A common method is to run MRI scans through brain segmentation pipelines, such as FreeSurfer \cite{fischl2012freesurfer}, to compute values measuring the anatomical structure of the brain, and enter those values into a statistical or machine learning model to classify the state of the brain \cite{gerardin2009multidimensional,plant2010automated,bhagwat2018modeling}. 

With the advent of big data and deep learning, convolutional neural network models have become increasingly popular for the task of image classification. While a variety of architectures and training methods have been proposed, the advantages and disadvantages of these design choices are not clear.

\cite{hon2017towards} used the pre-trained 2D VGG model \cite{simonyan2014very} and Inception V4 model \cite{szegedy2017inception} to train on a set of 6,400 slices from the axial view of 200 patients' MRI scans from the OASIS dataset to perform two-class (Normal vs. Alzheimer's) classification, where 32 slices are extracted from each patient's MRI scans based on the entropy of the image. 

\cite{jain2019convolutional} followed a similar approach using a 2D VGG model \cite{simonyan2014very} on the ADNI dataset to train on a set of 4,800 brain images augmented from 150 subjects' scans. For each subject's scan, they selected 32 slices based on the entropy of the slices to compose the dataset. They then proceed to shuffle the data and split it with a 4:1 training to testing ratio to perform the more complicated task of three-class classification for Normal, Mild Cognitive Impairment, and Alzheimer.

Other methods used 3D models, which are more computationally expensive but have more learning capacity for the three-dimensional MRI data.

\cite{payan2015predicting} pretrained a sparse autoencoder on randomly generated $5\times5\times5$ patches and used the weights as part of a 3D CNN to perform classification on the ADNI dataset, splitting randomly by MRI. \cite{hosseini2016alzheimer} employed a similar approach of using sparse autoencoder to pretrain on scans, but instead of randomly selecting patches from the training dataset, they used scans from the CADDementia dataset \cite{bron2015standardized}, and performed ten-fold cross-validation on the ADNI dataset.

\cite{khvostikov20183d} extracted regions of interest around the lobes of the hippocampus using atlas Automated Anatomical Labeling (AAL), augmented their data by applying random shifts of their data by up to two pixels and random Gaussian blur with $\sigma$ up to 1.2, and classified using a Siamese network on each of the regions of interest.

More recently, \cite{wang2018AD} trained a 3D DensetNet with ensemble voting on the ADNI dataset to achieve the state-of-the-art accuracy on three-class classification. They split by patients but treated MRIs of the same patient that are over three years apart as different patients, giving away some information from the training to testing process.

Table \ref{tab:different_paper} summarizes information on the data and evaluation approaches used by a few recent papers. 

Unlike most of the aforementioned papers that report high performance but do not explicitly mention their training and testing data split methodology, \cite{backstrom2018efficient} pointed out the problem of potentially giving away information from the training set into the testing set when splitting randomly by MRI scans, and showed that splitting MRI data by patient led to worse model performance. However, they only report two-class classification (Normal vs. Alzheimer's) in their results. They also only used a subset that is less than half the size of the MRIs available in ADNI, even though they reported empirical studies showing that \textit{the same model performing well on a small dataset can experience a significantly reduced performance on a large dataset}.

Our work differs in that we use as much of the data as available from ADNI and provide results in three-class classification (CN vs MCI vs AD). In addition, we introduce a method of splitting the data by visit history motivated by real-world use case that is new to the application in Alzheimer's disease brain MRI classification, and we compare the effects of the different splitting approaches in model performance.

\begin{table*}
\centering
\begin{tabular}{r|c|c|c|c}
\hline
  & \# MRIs & \# Patients & Data Splitting Approach & Accuracy \\
\hline \hline
\cite{jain2019convolutional}  & 150  & 150  &  8:2 train/test split, by augmented MRI slices & 95.7\% (2D) \\
\hline
\cite{farooq2017deep} &  355 &  149  & 3:1 train/test split, by augmented MRI slices & 98.8\% (2D) \\ \hline
\cite{payan2015predicting} & 100 & n/a & 8:1:1 train/val/test split, by patches from MRI & 89.5\% (3D) \\
\hline
\cite{hosseini2016alzheimer} &  210 & 210 &  10-fold cross validation, by patient & 94.8\% (3D) \\ \hline
\cite{khvostikov20183d} & 214  & 214  &  5:1 train/test split, by patient & 96.5\% (3D) \\
\hline
\cite{wang2018AD} &  833 &  624  &  7:3 train/test split, by patient \& MRI & 97.2\% (3D) \\ \hline
\end{tabular}
\caption{Performance and methodology of some of the state-of-the-art studies on three-class Alzheimer's Disease classification. The different splitting schemes and subsets of the ADNI dataset used in evaluation make it hard to interpret the results meaningfully.}
\label{tab:different_paper}
\end{table*}

\section{Methodology and Results}

In this section, we will briefly discuss the dataset we worked with, the preprocessing steps we took to produce our results, the model architectures we chose and training hyperparameters we used, as well as the results we obtained from the various methods of splitting the dataset.

\subsection{Data Acquisition and Preprocessing}
We use MRI data from ADNI because it is the largest publicly available dataset on Alzheimer's Disease, consisting of patients who experience mental decline getting their brain scanned and cognition assessed in longitudinal visits over the span of several years. 

We collected all of the scans that underwent \textit{Multiplanar Reconstruction (MPR)}, \textit{Gradwarp}, and \textit{B1 Correction} pre-processing from the dataset, discarding the replicated scans of the brain for the same patient during the same visit. We performed statistical parametric mapping (SPM) \cite{ashburner2005unified} using the T1--volume pipeline in the Clinica software platform (www.clinica.run), developed by the ARAMIS lab (www.aramislab.fr), as an additional pre-processing step on the images to spatially normalize them. 

In total, 2,731 MRIs from 657 patients are selected for our experiments. For the split by patient history, we used 2,074 scans for training and validation, and 657 scans for testing, where each scan in the testing set is from the last visit of the patient. For splitting by MRI, we used the same number of scans for all sets as the previous split by patient history. And finally, for splitting at the patient level, we had 2,074 scans from 484 patients in the training set and validation set, and 657 scans from 173 patients in the testing set. For consistency, we used the same number of scans in all subsets for all three splits. Our code repository is publicly available \footnote{https://github.com/Information-Fusion-Lab-Umass/alzheimers-cnn-study}, and it includes the patients ids and visits that we evaluated.

\subsection{Model Architecture and Training}
\label{sec:model_arch}

We compared 2D and 3D CNN architecture performances. For our 2D CNN architecture, we used ResNet18 \cite{he2016deep} that was pretrained on ImageNet, allowing the model to learn how to better extract low-level features from images. 

\begin{figure}[ht]
    \centering
    \includegraphics[width=0.7\linewidth]{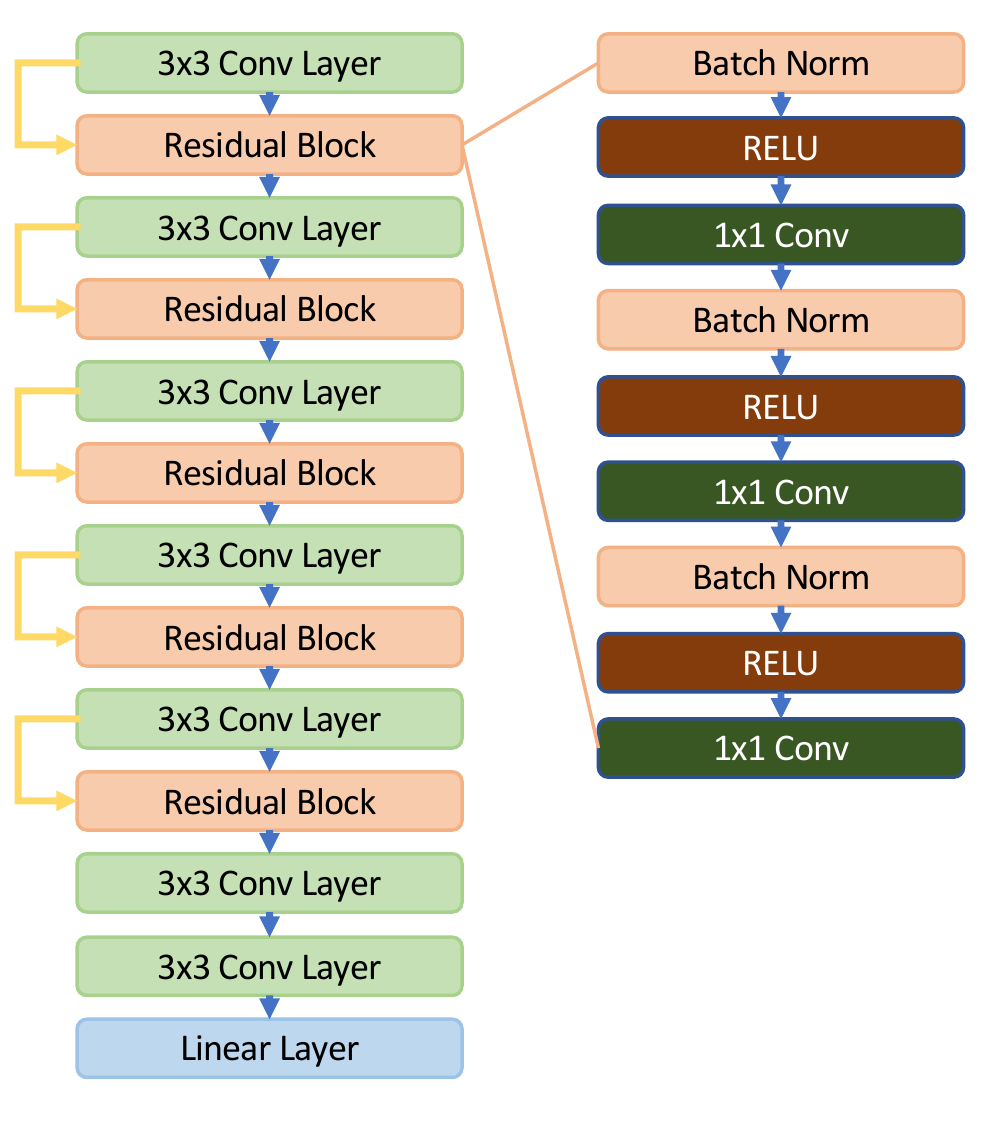}
    \caption{The architectures of our 3D CNN model and residual blocks. With the exception of the first and last convolution layers, which have a stride of 1, all other layers have a stride of 2 for down-sampling. The first convolution layers takes in a 1-channel image and outputs a 32-channel output.}
    \label{fig:network_architecture}
\end{figure}

For our 3D CNN architecture, we followed a residual network-based architectural design as well. We used the ``bottleneck" configuration, where the inner convolutional layer of each residual block contains half the number of filters, and we used the ``full pre-activation" layout for the residual blocks. See Figure \ref{fig:network_architecture} for more details.

For the 2D networks, we used a learning rate of 0.0001 and L2 regularization constant of 0.01.  For the 3D network, we used a learning rate of 0.001 and regularization constant of 0.0001. Both networks were trained for 36 epochs, with early stopping.

\begin{figure*}[!ht]
    \centering
        \centering
        \includegraphics[trim=200 864 200 893,width=0.6\linewidth]{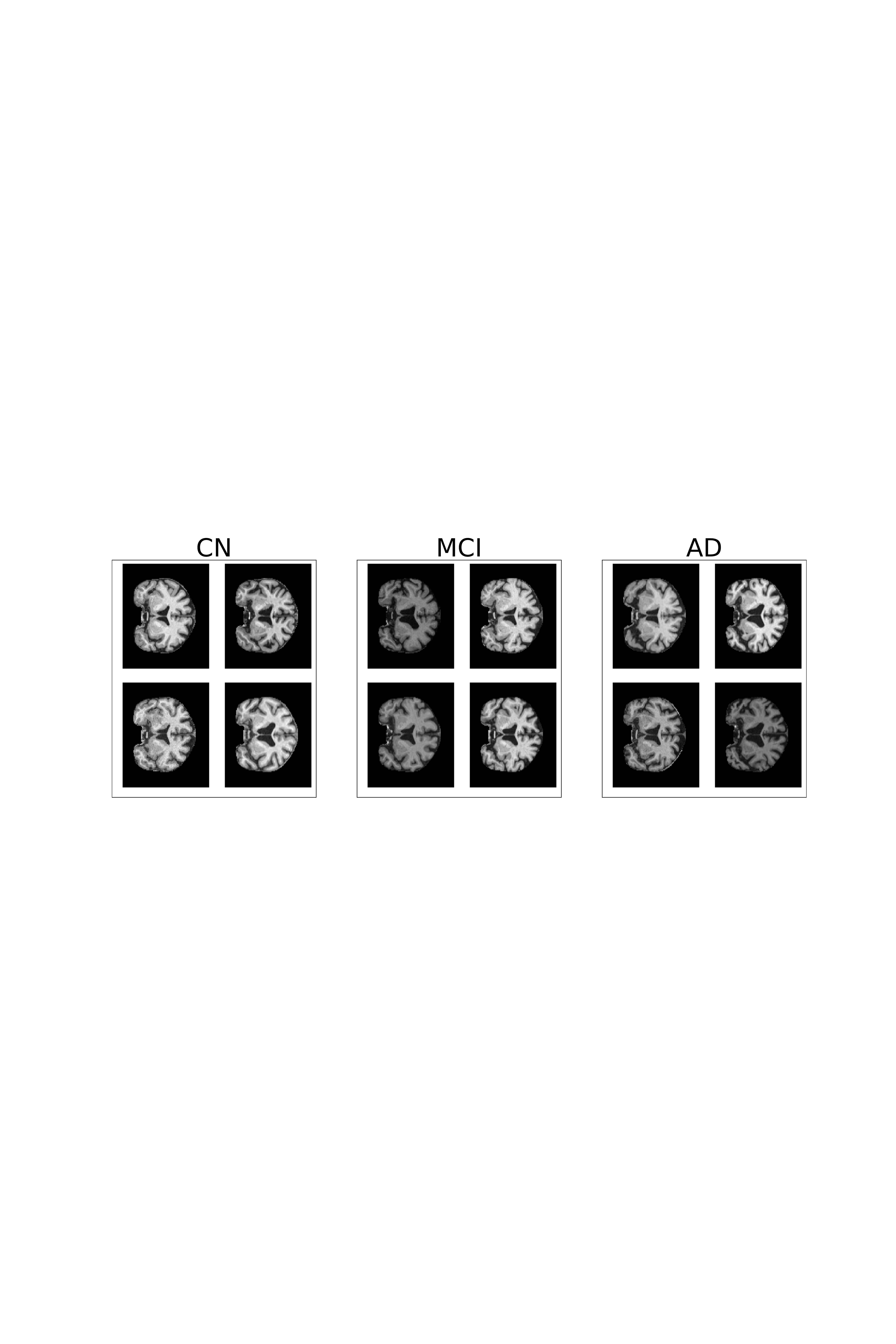}
       \caption{Comparison of the spatially normalized MRI scans of 4 subjects in each of the CN, MCI, and AD categories. To the human eye, distinguishing the difference across the disease stages is a difficult task.}
   \label{fig:mri_compare}
\end{figure*}

\subsection{Splitting Methodology and Results}
Our main goal is to investigate how differently the model performs under the following three scenarios: (1) random training and testing split across the brain MRIs, (2) training and testing split by patient ID, and (3) training and testing split based on visit history across the patients.

\begin{table}[H]
\centering
\begin{tabular}{c|l|c c}  
\hline
\multicolumn{2}{c}{Model} & Train Acc. & Test Acc. \\
\hline
& Split by MRI randomly & $99.2 \pm 0.7 \%$  & $83.7 \pm 1.1 \%$     \\
2D & Split by visit history & $99.0 \pm 0.5 \%$ & $81.2 \pm 0.5 \%$   \\
& Split by patients  & $98.8 \pm 0.6 \%$  & $51.7 \pm 1.2 \%$ \\
\hline
& Split by MRI randomly & $95.8 \pm 2.3 \%$ & $84.4 \pm 0.6 \%$     \\
3D & Split by visit history & $95.8 \pm 2.2 \%$  & $82.9 \pm 0.3 \%$      \\
& Split by patients  & $86.3 \pm 9.5\%$  & $52.4 \pm 1.8 \%$     \\
\hline
\end{tabular}
\caption{Our classification result of CNN models by different train/test splitting scheme, averaged over five runs.}
\label{tab:results101}
\end{table}

In Table \ref{tab:results101}, we present the three-way classification result of the models described in Section \ref{sec:model_arch}. In summary, when the training and testing sets were split by MRI scans randomly, the 2D and 3D models attained accuracy close to 84\%. When the training and test sets were split by visits, in which only the latest visit of each patient was set aside for testing, accuracy was slightly lower. However, accuracy dropped to around 50\% when the training and test set were split by patient.

We note that the other state-of-the-art 2D CNN architectures we tried (DenseNet, InceptionNet, VGGNet) performed similarly to ResNet, and the choice of view in the 2D slice (coronal, axial, sagittal) did not lead to significant differences in testing accuracy as long as the slices were chosen to be close to the center of the brain. We report results on the 88th slice of the coronal view for our 2D model. Our 3D model performs slightly better but is still limited in performance, suffering the same problem of over-fitting to information on individual patients instead of learning what generally differentiates a brain in the different stages of Alzheimer's Disease.

\begin{table}[H]
\centering
\begin{tabular}{c r|c|c|c}
\hline
& & \multicolumn{3}{c}{Ground Truth} \\

& & AD & MCI & CN \\
\hline
\multirow{3}{4em}{Prediction}   & AD & 96 & 59 & 17 \\
                                & MCI & 62 & 153 & 90 \\ 
                                & CN & 17 & 67 & 96 \\
\hline
\end{tabular}
\caption{Confusion matrix of the 3D CNN experiment on the test set, with data split by patient.}
\label{tab:results102}
\end{table}

\subsection{Analysis}

Analysis of the dataset shows that the frequency of disease stage transition for patients between any two consecutive visits is low in the ADNI dataset. There are only 152 transitions in the entire dataset, which contains 2,731 scans from 657 patients. 

We believe that due to the relatively few transition points in the dataset, the models are still able to achieve accuracy in the 80\% range for the splitting by visit experiments by repeating the diagnostic label of the previous visits. Out of the 152 total transitions we found across the whole dataset, only 52 of them happened for patients between the $n-1$\textsuperscript{th} visit and $n$\textsuperscript{th} visit. This suggests that the model is able to encode the structure of a patient's brain from the training set, in turn aiding its performance on the testing set.

Furthermore, we set up additional experiments where we trained on visits $t_{1} ... t_{n-1}$ from all patients, but only tested on the 52 patients that had a transition from the $n-1$\textsuperscript{th} to $n$\textsuperscript{th} visit. The classification accuracy on this experiment dropped to around 54\%, which suggests that the network was repeating information about the patient's brain structure instead of learning to be discriminative among the different stages of Alzheimer's Disease exhibited by a particular MRI scan.

\section{Discussion}

\subsection{Technical Challenges}

We summarize the main challenges in working with the ADNI dataset as follows:

1. \textbf{Lack of transitions in a patient's health status between consecutive visits.} There are only 152 transitions total out of the entire dataset of 2,731 images collected from patient visits. It is easy for the model to overfit and memorize the state of a patient at each visit instead of generalizing the key distinctions between the different stages of Alzheimer's Disease.

2. \textbf{Coarse-grained data labels.} The data labels are coarse-grained in nature so our classifier may become confused when trying to learn on cases when a patient's cognitive state may be borderline, such as being between MCI and AD. The confusion matrix in Table \ref{tab:results102} demonstrates this. 

3. \textbf{Difficulty in distinguishing the visual difference of a brain in the different Alzheimer's stages.} Human brains are distinct by nature, and the quality of MRI collections from different clinical settings add to the noise level of the data. In Figure \ref{fig:mri_compare}, we plotted out the brains of subjects in CN/MCI/AD, and show that the difference in anatomical structure from CN to MCI to AD is very subtle. 

4. \textbf{No clear baseline.} Many studies evaluated the performance of their models on different subsets of the ADNI dataset, making fair comparison a tricky task. In addition, studies that use a separate testing set do not report the subjects or scans that they used in their testing set, further complicating the comparison process.

We hope to keep these challenges in mind when designing future experiments and ultimately design models that can reliably classify brain MRIs with its true stage in Alzheimer's Disease progression, which is robust to visit number, lack of patient transitions, and minor fluctuations in scan quality.

\subsection{Insights and Future Work}

Many studies in Alzheimer's disease brain MRI classification do not take into account how the data should be properly split, putting into question the ability of the proposed models to generalize on unseen data. We fill this gap by providing detailed analysis of model performance across splitting schemes. Additionally, to our knowledge, none of the previous studies use all of the MRIs available in the ADNI dataset and do not present a clear explanation for this decision. To address the issue, we perform our experiments on all available data while also reporting the subjects used in the training and test split of all our experiments for reproducibility. 

In the future, we would like to explore utilizing the covariate data collected from patients to aid image feature extraction. Most of the studies we have come across do not use any covariate information collected from patients. The covariates, such as patient demographics and cognitive test scores, may be helpful for the classification task since they correlate with the disease stage of the patient. A scenario could be a multitask learning setup, where the model predicts the Mini-Mental State Examination (MMSE) and Alzheimer's Disease Assessment Scale (ADAS) cognitive scores in addition to the labels. We think this may be helpful in training the model because the cognitive test scores can provide finer-grained signal for the model, making the prediction more robust.

\clearpage

\bibliographystyle{named}
\bibliography{ijcai19}

\begin{thebibliography}{}

\bibitem[\protect\citeauthoryear{Ashburner and
  Friston}{2005}]{ashburner2005unified}
John Ashburner and Karl~J Friston.
\newblock Unified segmentation.
\newblock {\em Neuroimage}, 26(3):839--851, 2005.

\bibitem[\protect\citeauthoryear{Association and
  others}{2017}]{alzheimer2017facts}
Alzheimer's Association et~al.
\newblock 2017 alzheimer's disease facts and figures.
\newblock {\em Alzheimer's \& Dementia}, 13(4):325--373, 2017.

\bibitem[\protect\citeauthoryear{B{\"a}ckstr{\"o}m \bgroup \em et al.\egroup
  }{2018}]{backstrom2018efficient}
Karl B{\"a}ckstr{\"o}m, Mahmood Nazari, Irene Yu-Hua Gu, and Asgeir~Store
  Jakola.
\newblock An efficient 3d deep convolutional network for alzheimer's disease
  diagnosis using mr images.
\newblock In {\em 2018 IEEE 15th International Symposium on Biomedical Imaging
  (ISBI 2018)}, pages 149--153. IEEE, 2018.

\bibitem[\protect\citeauthoryear{Bhagwat \bgroup \em et al.\egroup
  }{2018}]{bhagwat2018modeling}
Nikhil Bhagwat, Joseph~D Viviano, Aristotle~N Voineskos, M~Mallar Chakravarty,
  Alzheimer’s Disease~Neuroimaging Initiative, et~al.
\newblock Modeling and prediction of clinical symptom trajectories in
  alzheimer’s disease using longitudinal data.
\newblock {\em PLoS computational biology}, 14(9):e1006376, 2018.

\bibitem[\protect\citeauthoryear{Bron \bgroup \em et al.\egroup
  }{2015}]{bron2015standardized}
Esther~E Bron, Marion Smits, Wiesje~M Van Der~Flier, Hugo Vrenken, Frederik
  Barkhof, Philip Scheltens, Janne~M Papma, Rebecca~ME Steketee,
  Carolina~M{\'e}ndez Orellana, Rozanna Meijboom, et~al.
\newblock Standardized evaluation of algorithms for computer-aided diagnosis of
  dementia based on structural mri: the caddementia challenge.
\newblock {\em NeuroImage}, 111:562--579, 2015.

\bibitem[\protect\citeauthoryear{Farooq \bgroup \em et al.\egroup
  }{2017}]{farooq2017deep}
Ammarah Farooq, SyedMuhammad Anwar, Muhammad Awais, and Saad Rehman.
\newblock A deep cnn based multi-class classification of alzheimer's disease
  using mri.
\newblock In {\em 2017 IEEE International Conference on Imaging systems and
  techniques (IST)}, pages 1--6. IEEE, 2017.

\bibitem[\protect\citeauthoryear{Fischl}{2012}]{fischl2012freesurfer}
Bruce Fischl.
\newblock Freesurfer.
\newblock {\em Neuroimage}, 62(2):774--781, 2012.

\bibitem[\protect\citeauthoryear{Gerardin \bgroup \em et al.\egroup
  }{2009}]{gerardin2009multidimensional}
Emilie Gerardin, Ga{\"e}l Ch{\'e}telat, Marie Chupin, R{\'e}mi Cuingnet,
  B{\'e}atrice Desgranges, Ho-Sung Kim, Marc Niethammer, Bruno Dubois,
  St{\'e}phane Leh{\'e}ricy, Line Garnero, et~al.
\newblock Multidimensional classification of hippocampal shape features
  discriminates alzheimer's disease and mild cognitive impairment from normal
  aging.
\newblock {\em Neuroimage}, 47(4):1476--1486, 2009.

\bibitem[\protect\citeauthoryear{He \bgroup \em et al.\egroup
  }{2016}]{he2016deep}
Kaiming He, Xiangyu Zhang, Shaoqing Ren, and Jian Sun.
\newblock Deep residual learning for image recognition.
\newblock In {\em Proceedings of the IEEE conference on computer vision and
  pattern recognition}, pages 770--778, 2016.

\bibitem[\protect\citeauthoryear{Hon and Khan}{2017}]{hon2017towards}
Marcia Hon and Naimul~Mefraz Khan.
\newblock Towards alzheimer's disease classification through transfer learning.
\newblock In {\em 2017 IEEE International Conference on Bioinformatics and
  Biomedicine (BIBM)}, pages 1166--1169. IEEE, 2017.

\bibitem[\protect\citeauthoryear{Hosseini-Asl \bgroup \em et al.\egroup
  }{2016}]{hosseini2016alzheimer}
Ehsan Hosseini-Asl, Georgy Gimel'farb, and Ayman El-Baz.
\newblock Alzheimer's disease diagnostics by a deeply supervised adaptable 3d
  convolutional network.
\newblock {\em arXiv preprint arXiv:1607.00556}, 2016.

\bibitem[\protect\citeauthoryear{Jain \bgroup \em et al.\egroup
  }{2019}]{jain2019convolutional}
Rachna Jain, Nikita Jain, Akshay Aggarwal, and D~Jude Hemanth.
\newblock Convolutional neural network based alzheimer’s disease
  classification from magnetic resonance brain images.
\newblock {\em Cognitive Systems Research}, 2019.

\bibitem[\protect\citeauthoryear{Khvostikov \bgroup \em et al.\egroup
  }{2018}]{khvostikov20183d}
Alexander Khvostikov, Karim Aderghal, Jenny Benois-Pineau, Andrey Krylov, and
  Gwenaelle Catheline.
\newblock 3d cnn-based classification using smri and md-dti images for
  alzheimer disease studies.
\newblock {\em arXiv preprint arXiv:1801.05968}, 2018.

\bibitem[\protect\citeauthoryear{Korolev \bgroup \em et al.\egroup
  }{2017}]{korolev2017residual}
Sergey Korolev, Amir Safiullin, Mikhail Belyaev, and Yulia Dodonova.
\newblock Residual and plain convolutional neural networks for 3d brain mri
  classification.
\newblock In {\em 2017 IEEE 14th International Symposium on Biomedical Imaging
  (ISBI 2017)}, pages 835--838. IEEE, 2017.

\bibitem[\protect\citeauthoryear{Payan and Montana}{2015}]{payan2015predicting}
Adrien Payan and Giovanni Montana.
\newblock Predicting alzheimer's disease: a neuroimaging study with 3d
  convolutional neural networks.
\newblock {\em arXiv preprint arXiv:1502.02506}, 2015.

\bibitem[\protect\citeauthoryear{Plant \bgroup \em et al.\egroup
  }{2010}]{plant2010automated}
Claudia Plant, Stefan~J Teipel, Annahita Oswald, Christian B{\"o}hm, Thomas
  Meindl, Janaina Mourao-Miranda, Arun~W Bokde, Harald Hampel, and Michael
  Ewers.
\newblock Automated detection of brain atrophy patterns based on mri for the
  prediction of alzheimer's disease.
\newblock {\em Neuroimage}, 50(1):162--174, 2010.

\bibitem[\protect\citeauthoryear{Simonyan and
  Zisserman}{2014}]{simonyan2014very}
Karen Simonyan and Andrew Zisserman.
\newblock Very deep convolutional networks for large-scale image recognition.
\newblock {\em arXiv preprint arXiv:1409.1556}, 2014.

\bibitem[\protect\citeauthoryear{Szegedy \bgroup \em et al.\egroup
  }{2017}]{szegedy2017inception}
Christian Szegedy, Sergey Ioffe, Vincent Vanhoucke, and Alexander~A Alemi.
\newblock Inception-v4, inception-resnet and the impact of residual connections
  on learning.
\newblock In {\em Thirty-First AAAI Conference on Artificial Intelligence},
  2017.

\bibitem[\protect\citeauthoryear{Wang \bgroup \em et al.\egroup
  }{2018}]{wang2018AD}
Shuqiang Wang, Hongfei Wang, Yanyan Shen, and Xiangyu Wang.
\newblock Automatic recognition of mild cognitive impairment and alzheimers
  disease using ensemble based 3d densely connected convolutional networks.
\newblock In {\em 2018 17th IEEE International Conference on Machine Learning
  and Applications (ICMLA)}, pages 517--523. IEEE, 2018.

\bibitem[\protect\citeauthoryear{Wu \bgroup \em et al.\egroup
  }{2018}]{wu2018discrimination}
Congling Wu, Shengwen Guo, Yanjia Hong, Benheng Xiao, Yupeng Wu, Qin Zhang,
  Alzheimer’s Disease~Neuroimaging Initiative, et~al.
\newblock Discrimination and conversion prediction of mild cognitive impairment
  using convolutional neural networks.
\newblock {\em Quantitative Imaging in Medicine and Surgery}, 8(10):992, 2018.

\end{thebibliography}

\end{document}